\documentclass[
	a4paper,
	10pt, 
	unnumberedsections,
	twoside,
]{LTJournalArticle}

\usepackage{graphicx}
\usepackage{subcaption}
\usepackage{amsmath}
\usepackage{url}
\usepackage{xcolor}
\usepackage{tikz}
\usetikzlibrary{positioning,fit,arrows.meta,shapes.multipart}

\footertext{\emph{RISC-V Summit Europe, Bologna, 8-12th June 2026}}

\title{FPGA Lifecycle Management for RISC-V Systems}

\author{Tianhai Liu\thanks{\small Corresponding author: \texttt{tianhai.liu@aicas.com} \\ This work is partially funded by the CHIPS-JU project, TRISTAN, grant number 101095947.}, James J. Hunt}
\date{
 \small aicas GmbH, Germany
}

\renewcommand{\maketitlehookd}{%
\begin{abstract}
\noindent
FPGA lifecycle management remains tied to proprietary toolchains and host architectures, leaving RISC-V without a vendor-neutral model for scalable bitstream deployment.
A host-agnostic control-plane architecture is presented that shifts lifecycle management to the operating-system layer by leveraging standard Linux capabilities, thereby decoupling deployment from specific ISAs and vendor stacks. This enables Linux-capable RISC-V processors to serve as control hosts in heterogeneous FPGA systems.
Prototyped on a Zynq-7000 SoC and generalizable to RISC-V platforms, the architecture provides a portable foundation for fleet-scale FPGA management.
\end{abstract}
}

\begin{document}

\maketitle

\section{Introduction}

Heterogeneous systems combine host processors with FPGA accelerators to meet performance and energy-efficiency requirements. The host executes control and system services, while compute-intensive workloads are offloaded to reconfigurable logic. Open instruction-set architectures such as RISC-V are gaining traction in heterogeneous SoCs that integrate processors with accelerators or FPGA fabric (e.g., Microchip PolarFire SoC, OpenHW CVA6).

However, FPGA bitstream deployment and lifecycle management remain largely platform-specific, typically tied to ARM- or x86-based systems, and tightly coupled to vendor ecosystems such as AMD/Xilinx (Vivado/Vitis), Intel FPGA (Quartus/Platform Designer), and cloud-specific infrastructures like AWS EC2 F1. This tight coupling results in limited portability and vendor lock-in. Moreover, many development workflows assume manual updates, physical access, and reboot-based reconfiguration, which do not scale.
As a consequence, the RISC-V ecosystem currently lacks a vendor-neutral architectural model for fleet-oriented FPGA lifecycle management that aligns with its \emph{openness} and \emph{modularity} principles.

This paper presents a \emph{host-agnostic control-plane architecture} for remote FPGA bitstream deployment and lifecycle management at fleet scale. The design specifies its requirements at the operating-system level and relies solely on standard Linux capabilities, including driver interfaces, filesystem access, process isolation, and secure communication. By decoupling lifecycle management from specific ISAs and vendor toolchains, the architecture can be realized on ARM-, x86-, or RISC-V-based systems without modification. This enables open RISC-V platforms to participate in mature, CI/CD-integrated FPGA management workflows without dependence on proprietary host infrastructures.


\section{Architecture}

The architecture is based on a cloud-based manager for devices.  A realtime OSGi~\footnote{Richard Hall, Karl Pauls, Stuart McCulloch, and David Savage. 2011. OSGi in Action: Creating Modular Applications in Java (1st. ed.). Manning Publications Co., USA.} Agent runs on each device and manages the FPGA software.  The key feature is the extension of the OSGi lifecycle and version management to artifacts outside the VM, in this case FPGA bitstreams.  This host-agnostic control-plane architecture for remote FPGA deployment and lifecycle management is divided into two conceptual planes: a \emph{control plane} and a \emph{reconfigurable acceleration plane}.

The \emph{control plane} runs on a Linux-capable host processor. It is responsible for the following activities.

\noindent
\textbf{Bitstream configuration via driver interface} \\
    The control plane triggers FPGA reconfiguration via a standard Linux device driver (e.g., by writing a bitstream to \texttt{/dev/fpga0}), thereby decoupling the deployment logic from vendor-specific configuration utilities.

\noindent\textbf{Version management and bitstream storage} \\
Bitstreams are stored in a versioned cloud repository integrated with CI/CD pipelines, including metadata on hardware compatibility and software dependencies. Remote devices fetch approved versions on demand, supporting staged deployment and rollback.

\noindent\textbf{Deployment orchestration with dependencies} \\
    The system ensures that required kernel modules, user-space services, or hardware dependencies are present before activating a new bitstream, enabling coordinated updates across FPGA logic and host software.

\noindent\textbf{Isolation and modular service control} \\
    FPGA-backed services are executed in isolated processes or containers, allowing controlled start/stop/restart operations without affecting unrelated system components.

\noindent\textbf{Secure communication with the backend} \\
    Bitstreams and lifecycle commands are retrieved from a cloud backend via authenticated and encrypted channels (e.g., HTTPS with mutual TLS), ensuring integrity and provenance verification.


The \emph{acceleration plane} comprises the FPGA fabric that hosts one or more hardware accelerators.
Bitstreams are treated as versioned deployment artifacts that can be dynamically loaded/unloaded at runtime.

The architecture relies on standard operating system capabilities and is therefore independent of a specific instruction set architecture.
The separation between the control plane and the acceleration plane ensures that reconfiguration of hardware components does not require a full system reboot and does not affect unrelated software services.

\section{Workflow}

\begin{figure}[tb]
  \centering
  \includegraphics[width=\linewidth]{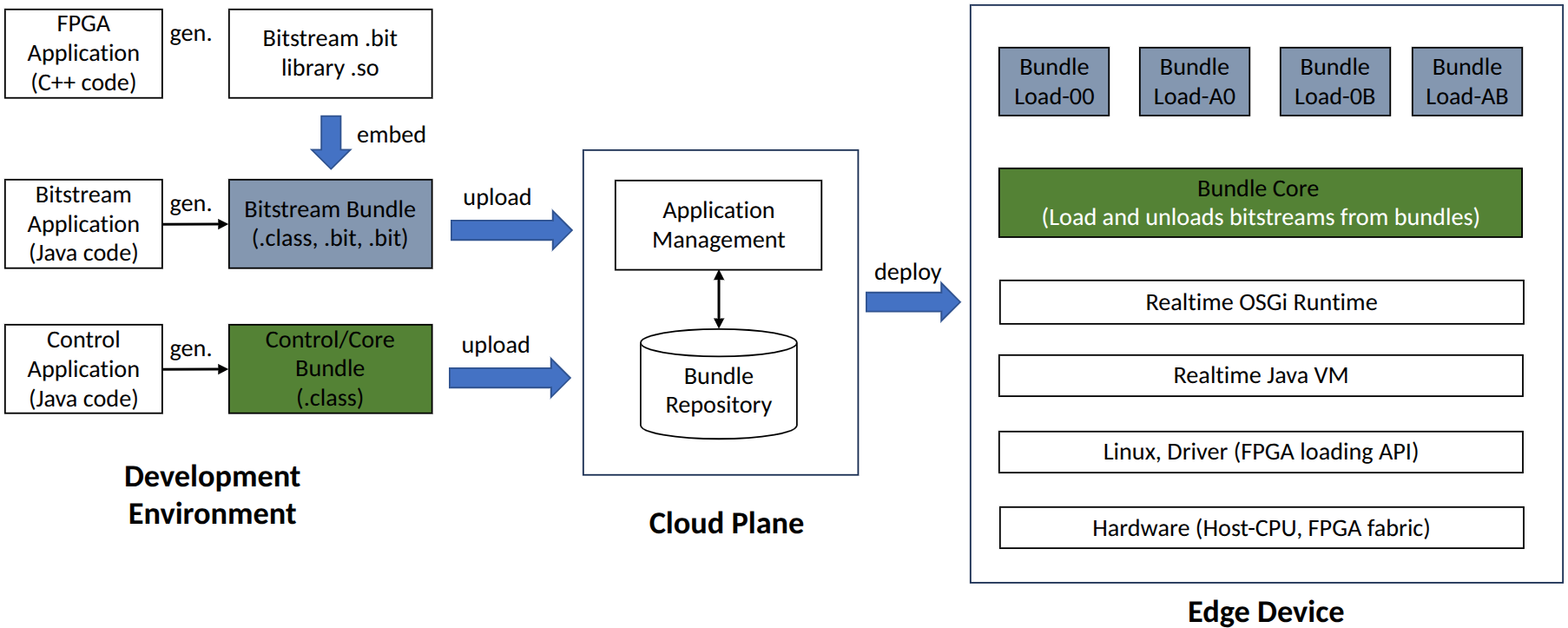}
  \caption{Architectural overview of the workflow.}
  \label{fig:workflow}
\end{figure}

Figure~\ref{fig:workflow} depicts the deployment and lifecycle management process.

\begin{enumerate}[leftmargin=*]
    \item \textbf{Artifact Packaging.}
    FPGA applications generate a bitstream (\texttt{.bit}) and associated native libraries (\texttt{.so}), which are packaged together with Java components into versioned deployment OSGi bundles.
    Control logic is packaged separately to preserve modularity between acceleration and orchestration.

    \item \textbf{Repository Integration and Validation.} 
    Bundles are uploaded to a central repository, where automated multi-stage validation pipelines perform consistency checks, integration testing, and policy verification before approval.

    \item \textbf{Remote Deployment.} 
    Approved artifacts are distributed by CloudPlane to selected edge devices via the CloudPlane client bundle through an encrypted communication channel.
    Deployment decisions can be version-based, environment-specific, or staged across device groups.

    \item \textbf{Atomic Reconfiguration.} 
    On the edge device, a dedicated loader/core bundle on control-plane invokes the FPGA driver interface to perform atomic bitstream reconfiguration.
    The host processor orchestrates the transition without requiring a full system reboot or restarting unrelated services.

    \item \textbf{Runtime Isolation.}
    The realtime OSGi runtime ensures that unrelated services remain active during updates, confining reconfiguration effects to the targeted acceleration component.

    \item \textbf{Rollback and Version Control.}
    If validation or runtime checks fail, the OSGi runtime supports controlled rollback to a previously deployed version.
\end{enumerate}

This workflow enables fleet-scale lifecycle management while preserving modularity, runtime isolation, and ISA independence at the architectural level.

\section{Implementation}

The proposed architecture has been implemented and demonstrated on an AMD Zynq-7000 SoC that integrates a dual-core ARM host processor and FPGA fabric on the same die.
The system runs PetaLinux, with a realtime OSGi runtime~\footnote{\url{https://www.aicas.com/products-services/jamaicaams/}} and a CloudPlane~\footnote{\url{https://www.aicas.com/products-services/aicas-edge-device-portal/}} client bundle providing modular service control.

The demonstrator implements multiple matrix-multiplication variants~\footnote{\url{https://github.com/Xilinx/SDAccel_Examples.git}} to illustrate dynamic FPGA lifecycle management. Several bitstream configurations and corresponding host-side shared libraries were generated in Vivado to enable selective acceleration of different computation modules.
A public demonstration video is available at \url{https://doi.org/10.5281/zenodo.18670854}.

In general, the architecture is hardware-agnostic and compatible with the following integration patterns.

\begin{itemize}[leftmargin=*]
    \item SoCs integrating host CPUs and FPGA fabric on the same die, such as AMD Zynq UltraScale+ MPSoC or Microchip PolarFire SoC.

    \item FPGA platforms partitioned into host-CPU and accelerator regions, where a Linux-capable soft-core (e.g., OpenHW CVA6) controls accelerator modules within the same fabric.

    \item Discrete FPGA accelerators connected via PCIe or similar interconnects, such as AMD Alveo cards or Amazon EC2 F1 instances.
\end{itemize}

\section{Conclusion}
This work presents a host-agnostic control-plane architecture for fleet-scale FPGA bitstream deployment and lifecycle management on heterogeneous edge platforms. Defining lifecycle requirements at the operating-system layer decouples FPGA management from specific processor vendors and proprietary toolchains.

The abstraction improves cross-platform portability and enables Linux-capable RISC-V systems to serve as control processors in FPGA-accelerated environments. More broadly, it supports standardized lifecycle management for heterogeneous edge infrastructures.

\printbibliography

\end{document}